# Energy management strategy for an optimum control of a standalone photovoltaic-batteries water pumping system for agriculture applications

BENZAOUIA Mohammed [1] [*], HAJJI Bekkay [1], ABDELHAMID Rabhi [2], MELLIT Adel [4], BENSLIMANE Anas [1] and ANNE Migan Dubois [3]

[1] Renewable Energy, Embedded System and Data Processing Laboratory, National School of Applied Sciences, Mohamed First University, Oujda, Morocco.
[2] Laboratory of Modelization, Information and Systems University of Picardie Jules Verne 3 rue Saint Leu - 80039 Amiens Cedex 1 – France.
[3] GeePs (Group of electrical engineering – Paris), UMR CNRS 8507, CentraleSupéle.
[4] Renewable Energy Laboratory, Faculty of Sciences and Technology, Jijel University, Algeria.

**Abstract.** Pumping water using multiple energy sources is the ideal solution for supplying of potable water in isolated or arid areas. In this paper, an effective control and energy management strategy for a stand-alone photovoltaic-batteries water pumping system for agriculture applications is presented. The system is composed of photovoltaic solar panels as primary energy sources, and Lead-Acid batteries as seconder energy sources to supply the brushless DC motor and the centrifugal pump. The energy management strategy uses an intelligent algorithm to satisfy the energy demanded by the motor, also to maintain the state-of-charge of the battery between safe margins in order to eliminate the full discharge and the destruction of the batteries. Drift is a major problem in photovoltaic systems; this phenomenon occurs when the solar irradiation changes rapidly. Classical MPPT algorithms do not solve this problem, for this reason a modified P&O has been implemented, the obtained results shown the efficiency of the algorithm compared to the conventional P&O. Computer simulation results confirm the effectiveness of the proposed energy management algorithm under random meteorological conditions.

## 1 Introduction

Water pumping is usually depends on conventional resources or a diesel generator . The use of these fossil energy resources (diesel generator, propane...) not only requires expensive fuels, but also creates noise and air pollution and high maintenance costs [1]. Combining several renewable energy sources such photovoltaic panels, and wind turbine with batteries and supercapacitors has received considerable attention recently, especially for pumping water in isolated areas where there is no grid power [2] and [3].

These systems are environment friendly, require low maintenance with no fuel cost, and provide a continuous energy whatever the variation of the load or the weather condition [4].

These hybrid systems required a management strategy in order to satisfy the load demand and to manage the power flow while ensuring efficient operation of the different energy systems. Various configurations and power management algorithms have been proposed in scientific research, in [5] the system consist of photovoltaic (PV) source, batteries storage, and an algorithm of power management, the system is used for electrification and water pumping. In [6] the authors propose an algorithm based on fuzzy logic controller to manage a hybrid system composed wind/photovoltaic/diesel with storage battery.

In the present work, photovoltaic panels and batteries with an intelligent power management control is applied to pump a constant flow of water under different conditions (Solar irradiation and state of charge). In order to ensure maximum extraction of power from the photovoltaic panels a Modified Perturb and Observe (MP&O) have been used. The brushless DC motor is controlled by electronic commutation.

## 2     Presentation & Modeling of studied system

The structure of the studied system is shown in Fig.1; it contains a photovoltaic generator connected to a DC-DC converter, which is controlled by the Modified P&O algorithm in order to extract the maximum power from the PV generators. The system also contains batteries connected to a bidirectional converter to allow the current to flow in both directions, so the batteries are charged and discharged according to the user's energy demand and the methodological conditions. An inverter is used as an intermediary between the two preceding components and the motor-pump. The control of the motor is done through the electronic commutation that allows to operate the motor in the nominal capacities. The system is managed with an intelligent algorithm, which depends on the state of the three switches ($K_1, K_2$ and $K_3$). The design of studied system is elaborated in following sections.

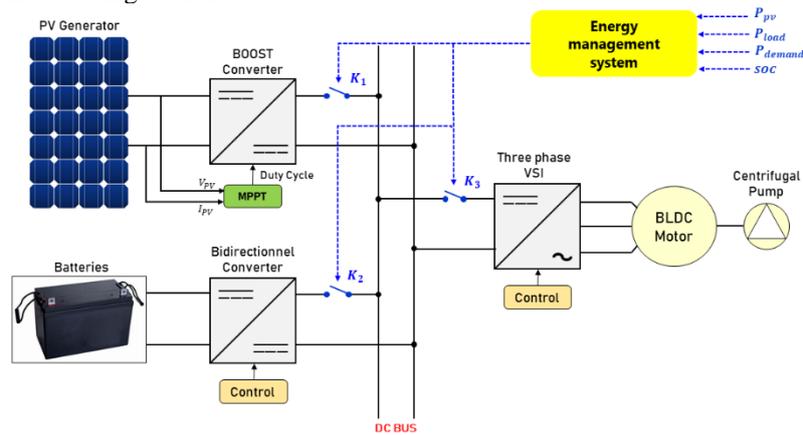

**Fig. 1.** Stand-alone water pumping system description.

### 2.1 Photovoltaic panels modeling

Several mathematical models of photovoltaic generators were developed to describe their nonlinear behavior and their operation. In this work, the following model (Fig.2) is chosen [7] [8] and [9]. The PV module parameters used in this work are shown in Table 1.

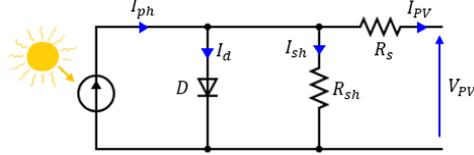

**Fig. 2.** PV cell equivalent circuit.

The output current ($I_{pv}$) of the photovoltaic cell under standard operating conditions ($1000\ w/m^2, 25°C$) is given by the following Eq. 1. and Eq. 2.

$$I_{PV} = I_{ph} - I_d - I_{sh} \quad (1)$$

$$I_{PV} = I_{ph} - I_s e^{(\frac{q(V_{PV}+ I.R_s)}{nkT})-1} - \frac{(V_{PV} + I.R_s)}{R_{sh}} \quad (2)$$

Where: $I_s$ is the saturation current, $q$ is the electron charge ($1,6.10^{-19}$(C)), $k$ is Boltzmann constant ($1,38.10^{-23}(J.K^{-1})$), $n$ is the diode ideality factor, $T$ is the PV cell temperature ($K$).

In this simulation, twenty four (24) panels has been used. The simulated $I$–$V$ and $P$–$V$ characteristics (Fig. 3. and Fig. 4.) present the effect of irradiation and temperature of the behavior of the adopted PV generators.

**Table 1.** The PV module parameters.

| Parameters | Values |
| --- | --- |
| Numbers of cells in a module, $N_c$ | 36 |
| Open circuit voltage, $V_{OC}$ | 21.8 V |
| Short circuit current, $I_{SC}$ | 7.24 A |
| Maximum voltage at MPP, $V_{MPP}$ | 17.2 V |
| Maximum current at MPP, $I_{MPP}$ | 6.69 A |
| Maximum Power at MPP, $P_{MPP}$ | 115 W |

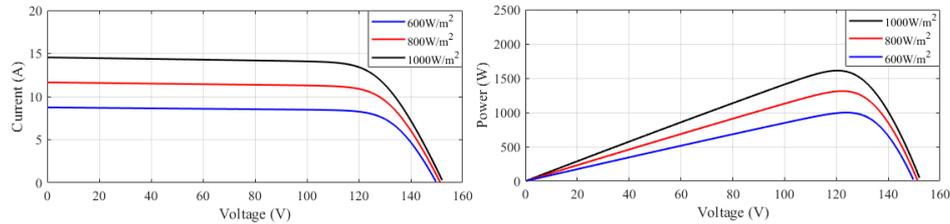

**Fig. 3.** *P–V* and *I-V* characteristics for different values of solar irradiation.

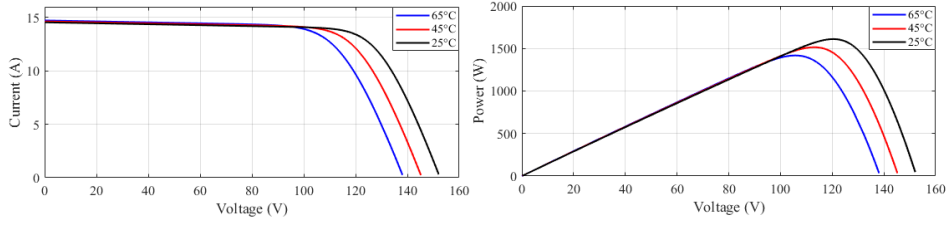

**Fig. 4.** *P–V* and *I-V* characteristics for different values of temperture.

### 2.2  Storage modeling

The purpose of using the batteries is to store the excess power and satisfy the load demand in bad weather conditions or in night periods. Many varieties model of batteries exist in the literature. In this work, the proposed model is shown in Fig. 5. It contains two electrical element: a voltage source ($E_B$) and an internal resistance ($R_i$) [10], [11], [12], and [13].

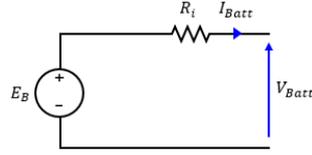

**Fig. 5.** Electrical model of the battery.

The adopted model of battery is defined by the following equations:
$$V_{Batt} = E_B \pm I_{Batt}\, R_i \tag{3}$$
Where: $E_B$ is the voltage source, $R_i$ the internal resistance and $I_{Batt}$ is the current of the battery.

The battery capacity $C_{Batt}$ is given by [14]:
$$C_{Batt} = \frac{E_d \cdot N_d}{V_{Batt} \cdot DOD \cdot \eta_{Batt}} \tag{4}$$
Where: $E_d$ is the daily electrical energy required by the load (Motor-pump), $N_d$ number of autonomy days, $V_{Batt}$ is the battery voltage, $DOD$ is the depth of discharge and $\eta_{Batt}$ is the battery performance.

The state of charge $SOC$ depends on the current of the battery ($I_{Batt}$). If it is positive (battery discharge) then the state will decrease and discharge if negative (battery charging) then the charge will increase [15]. This is given by the following equation:
$$SOC\,(t) = SOC(t-1) - \int_{t-1}^{t} \frac{I_{Batt}}{C_{Batt}}\, dt \tag{5}$$
Where: $C_{Batt}$ is the capacity of battery and $I_{Batt}$ represents the current that flow to or from the battery.

After sizing calculations, ten (10) batteries of 12 Volts, with a capacity unit $C_{Batt}$ =50Ah, have been considered.

## 3 Maximum power point tracking (MPPT)

Various algorithm were developed in order to extract the maximum power from photovoltaic panels. Perturb & Observe (P&O), Incremental conductance (Inc.) and Fuzzy logic (FL) are the most algorithm used in research [16] and [17]. In this work, a Modified Perturb & Observe (MP&O) is implemented to avoid the failure of P&O algorithm under fast changes of solar irradiation (Drift phenomenon) [18].

### 3.1 Perturb & Observe (P&O)

The P&O algorithm is widely used to extract the maximum power from *PV* panels. The simple structure and small measured parameters are among the reasons why this algorithm is chosen [19] and [20]. The entire P&O algorithm is given in Fig. 6.

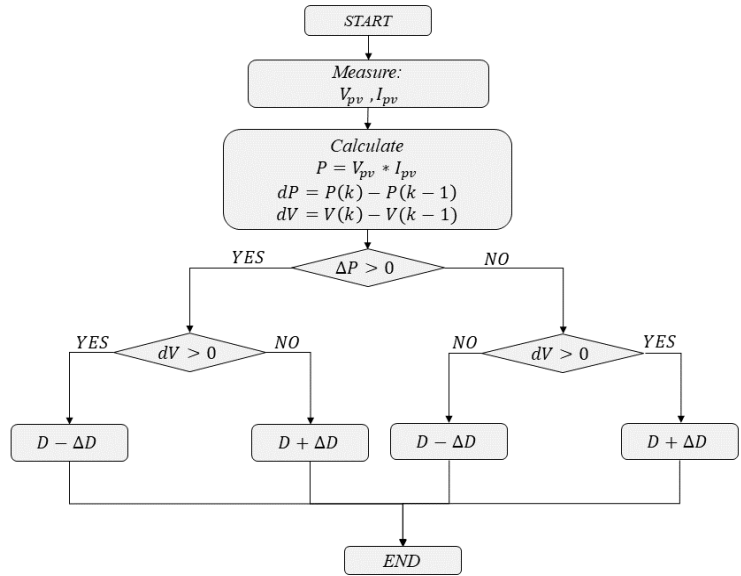

**Fig. 6.** Flowchart of the conventional P&O MPPT algorithm.

### 3.2 Modified Perturb & Observe (P&O)

The P&O technique algorithm fails under rapidly changing of solar irradiation [21], the P&O algorithm confused or taken improper decision at changing primary step of duty-cycle [22]. A modified P&O technique is proposed to avoid the drift problem by incorporating the information of change in current ($dI$) in the decision process in addition to change in power ($dP$) and change in voltage ($dV$). The steps of the algorithm are presented in Fig. 7 [23]. The obtained results (Fig. 8) show clearly the efficiency of the Modified Perturb & Observe (MP&O) algorithm compared to conventional P&O.

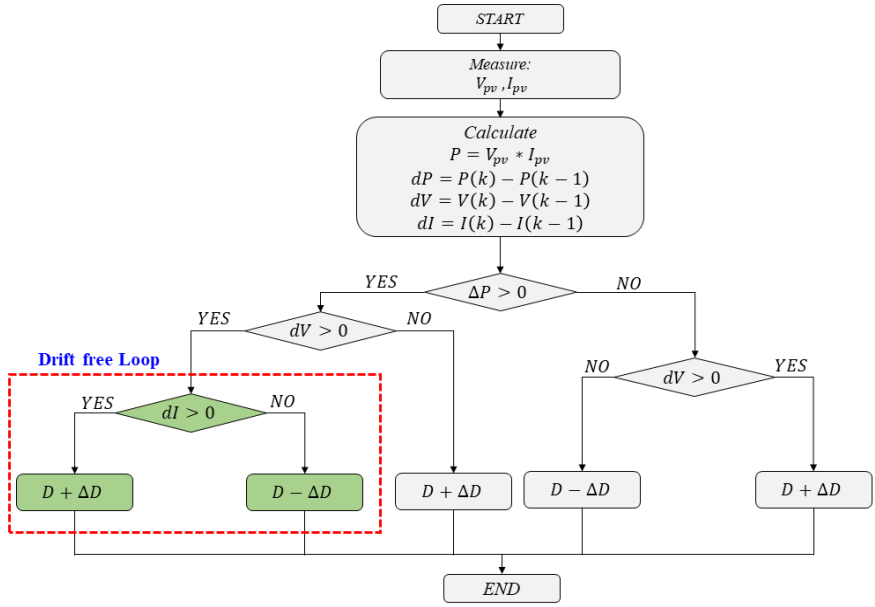

**Fig. 7.** Flowchart of the drift-free modified P&O MPPT algorithm.

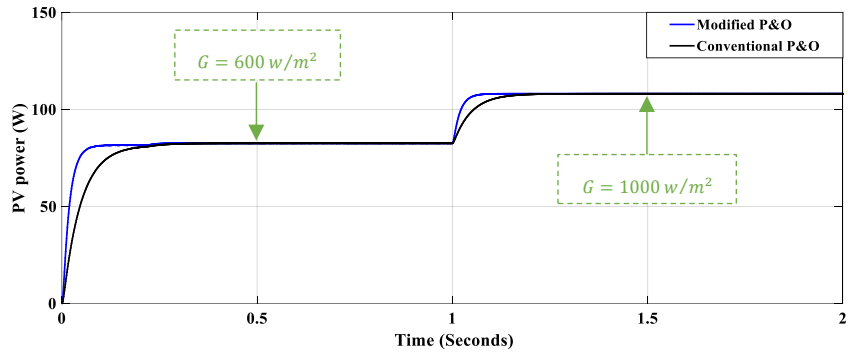

**Fig. 8.** Comparison between conventional & modification of P&O algorithm for one panel.

## 4  Energy management strategy

The power management for 48h is illustrated in the flowchart of Fig. 9. The strategy is based on the evaluation of daily demand of water ($P_{demand}$) by the user and on the availability of the power of solar ($P_{PV}$). The algorithm consists in generating three control signals $K_1, K_2$ and $K_3$ starting from four inputs: power of PV generators ($P_{PV}$), the power demand ($P_{demand}$) which is given by user, the measured power ($P_{Load}$) and the battery state of charge ($SOC$). For the beginning, it is necessary to set the maximum and minimum limits of the batteries $SOC$. In this work, the minimum $SOC_{min}$ is fixed to 20% and the maximum $SOC_{max}$ to 80%.

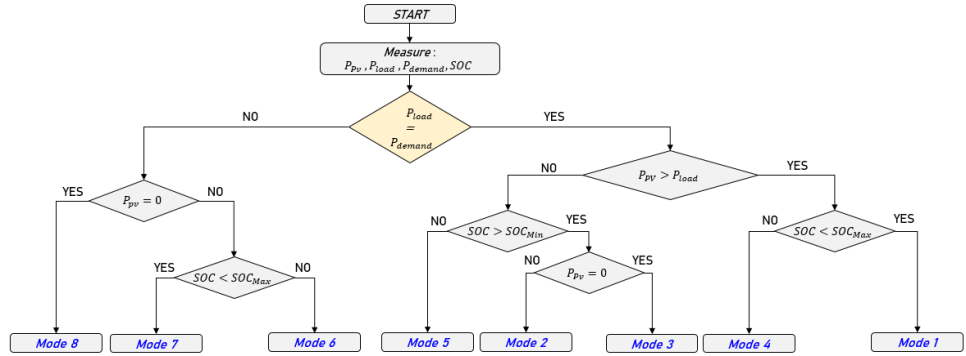

**Fig. 9.** Flowchart of energy management strategy.

The energy management strategy is described below:

➢ **If $P_{demand} = P_{Load}$ :**

**Mode 1 :** This mode of operation occurus during the following case : if the power generated by PV panels is greater than the power required by the load and the state of charge is less than $SOC_{max}$, the excess power is stored by the batteries.
**Mode 2 :** This mode of operation occurus during the following case : if the power generated by PV panels eqaul to the power demande and the $SOC = SOC_{max}$, so the disconnection of the batteries is necessary in order to protect them.
**Mode 3 :** This mode of operation occurus during the following case : if the power generated by PV panels equal zero and the $SOC > SOC_{min}$, this mode may accur during the night and the begining and end of the day, the batteries cover the power demanded by the load until the intervention of the PV generators.
**Mode 4 :** This mode of operation occurus during the following case : if the power generated by PV panels is less than the power demanded by the load, The batteries are used to add the necessair power as long as the state of charge is not less than the $SOC_{min}$.
**Mode 5 :** This mode of operation occurus during the following case : if the power generated by PV panels equal zero and the $SOC$ is less than $SOC_{min}$, The load is disconnected.

➢ **If $P_{demand} = 0$ :**

**Mode 6 :** This mode of operation occurus during the following case : the power of PV generators is $P_{PV} > 0$, in this case this power is stored in batteries, as long as the state of charge of the batteries does not reach maximum ($SOC < SOC_{max}$).

**Mode 7 :** This mode of operation occurus during the following case : the PV generators produce power, and the state of charge of the batteries is max in this case the batteries are diconnected.

**Mode 8 :** This mode of operation occurus during the following case : there is no load demand and the PV generators do not produce power so the system is disconnected.

## 5  Simulation & Results

In the studied system, the renewable PV power is taken as primary source, while the batteries are used as a backup and storage system. The priority is given to the PV generators, not only to exploit the entire renewable energy but also to increase the life cycle of the battery.

To test and verify the effectiveness of the energy management strategy applied to the studied system (Fig.1). The simulation under *MATLAB*/Simulink over a period of two different days has been performed.

Fig. 10 and Fig. 11 present the solar irradiation and temperature profiles, in the 1st day (00h-24h) the maximum irradiation value reaches $1000 \, w/m^2$, while in the 2nd day (24h-48h) it reaches $600 \, w/m^2$.

Depending on the variations in irradiation and temperature, the power profile produced by the PV generators is given by Fig. 12.

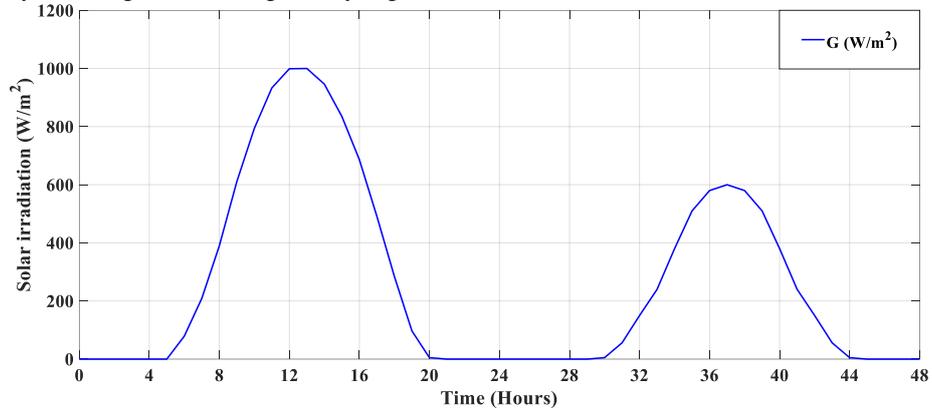

**Fig. 10.** Profile of the solar irradiation.

Fig. 13 shows the different powers supplied by the sources (PV generators and batteries), in addition to the load profile desired by the user, this profile ($P_{load}$) has been chosen in different periods (night, day, sunrise and sunset). Based on these results, we can differentiate four cases, which are given as follows:

**Case 1:** During the intervals [0h, 3h] and [23h, 26h], the load power demand (Motor-pump) is 1320W, the deficient power is only supplied by the batteries because $P_{PV} = 0$ and the state of charge ($SOC$) of batteries is not at the minimal value.

**Case 2:** In intervals [5h, 6h], [10h, 12h], [14h, 20h], [30h ,33h] and [41h, 44h] , the photovoltaic generators produce Power ($P_{PV} > 0$) so the batteries are in charge mode since there is no demand of power ($P_{Load} = 0$), and the $SOC_{min} < SOC < SOC_{max}$.

**Case 3:** In intervals [6h, 10h] and [33h ,41h], The load requires a power of 1320W but the photovoltaic generators do not produce sufficient power, the compensation of power is done by the batteries taking in consideration that the $SOC$ is not under $SOC_{min}$, ($SOC < SOC_{min}$).

**Case 4:** In interval [12h, 14h], there is a surplus of power produced by PV generators, and the load requires a power of 1320W, so the excess of power is stored in the batteries as long the $SOC < SOC_{max}$.

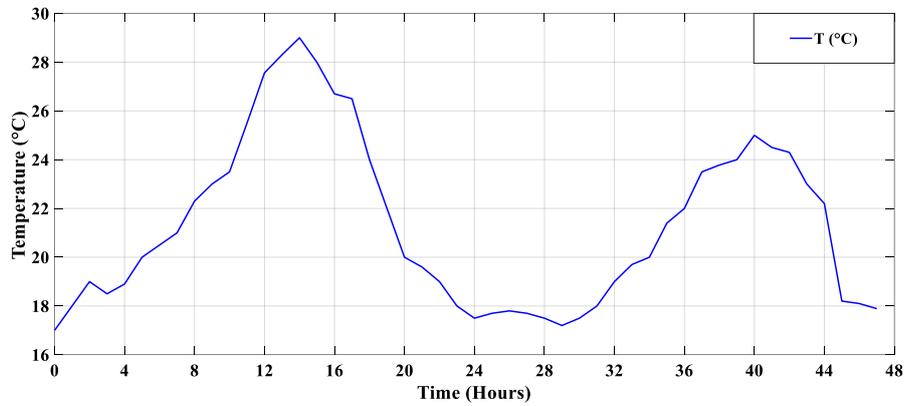

**Fig. 11.** Profile of ambient temperature.

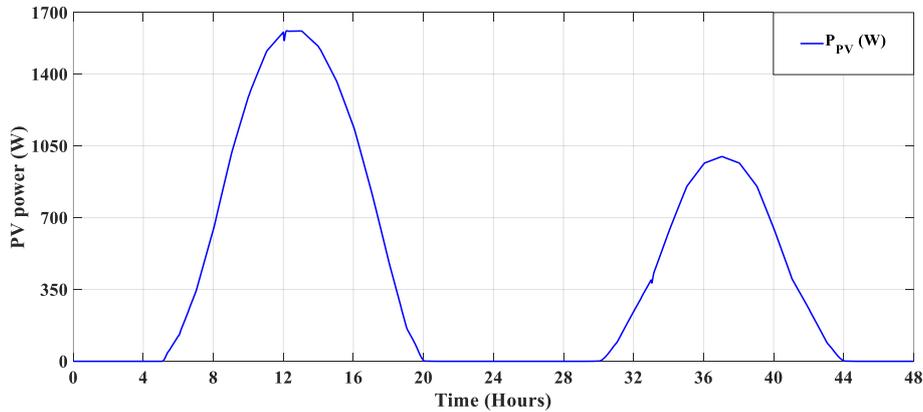

**Fig. 12.** Power of PV generator.

The evolution of the state of charge ($SOC$) is given in Fig. 14, the ($SOC$) varies between the minimum ($SOC_{min} = 20\%$) and the maximum ($SOC_{max} = 80\%$), therefore the batteries are protected against complete discharge and overcharging.

The objective was to pump a fixed water flow over two successive days and in different periods. Fig. 15 shows clearly that the pumped flow is fixed at 30 $m^3/h$.

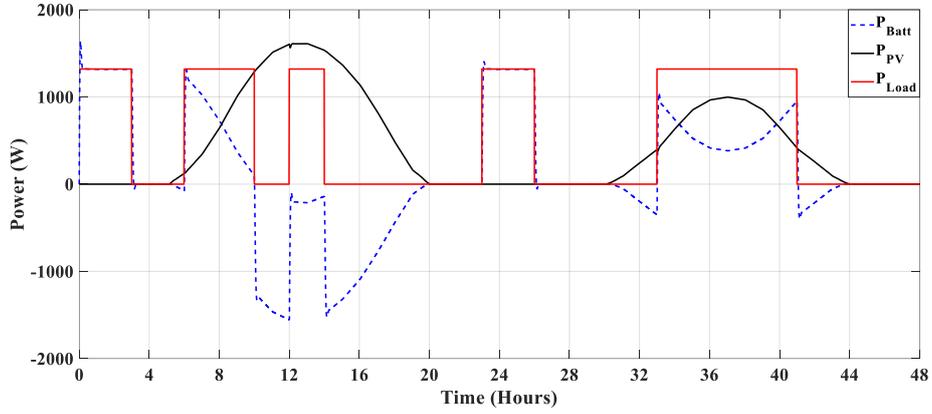

**Fig. 13.** Power profiles ($P_{Batt}, P_{PV}$ and $P_{Load}$).

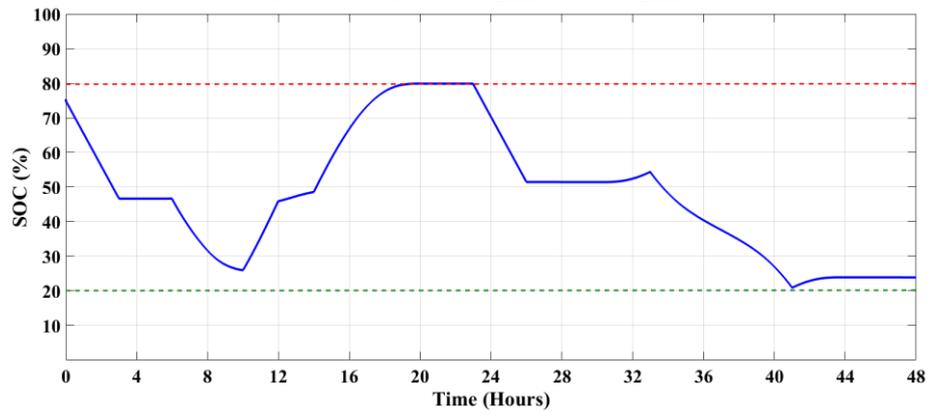

**Fig. 14.** State of charge of batteries.

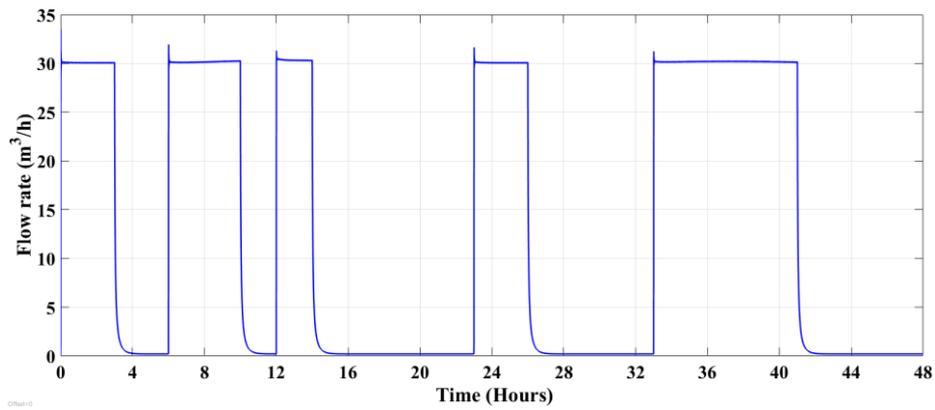

**Fig. 15.** Water flow.

## 6   Conclusion

In this paper, the control and management of a photovoltaic system with storage batteries for agricultural applications has been presented. The system configuration, the MPPT algorithm and the proposed management strategy were simulated in MATLAB/*Simulink* environment. The obtained results show the effectiveness and the reliability of the system during two successive days under different weather conditions. The proposed management strategy allow to ensure optimal operation of the system without human intervention, which makes the system more intelligent.